\documentclass{elsarticle}
\usepackage{epsf}

\title{SuperNOVA: a novel algorithm for graph automorphism calculations}
\author{Russell K. Standish}
\address{Mathematics and Statistics\\The University
  of New South Wales}
\ead{hpcoder@hpcoders.com.au}

\newcommand{\EcoLab}{{\sffamily\slshape
    \mbox{\raisebox{.5ex}{Eco}\hspace{-.4em}{\makebox[.5em]{L}ab}}}}

\begin{document}
\begin{abstract}
  The graph isomorphism problem is of practical importance, as well as
  being a theoretical curiosity in computational complexity theory in
  that it is not known whether it is $NP$-complete or $P$. However,
  for many graphs, the problem is tractable, and related to the
  problem of finding the automorphism group of the graph. Perhaps the
  most well known state-of-the art implementation for finding the
  automorphism group is Nauty. However, Nauty is particularly
  susceptible to poor performance on star configurations, where the
  spokes of the star are isomorphic with each other. In this work, I
  present an algorithm that explodes these star configurations,
  reducing the problem to a sequence of simpler automorphism group
  calculations.
\end{abstract}

\maketitle

\section{Introduction}

Given two graphs $g_1=\{V,E_1\}$ and $g_2=\{V,E_2\}$ where $V$ is a
set of labelled vertices, and $E_{\{1,2\}}\subset V\times V$ are sets
of edges, the {\em graph isomorphism} problem is the problem of
finding a permutation $\sigma: V\longrightarrow V$ of the vertices
such that $\forall (i,j)\in E_1, (\sigma(i),\sigma(j))\in E_2$. The
map $\sigma$ is known as an {\em isomorphism}.  The graph isomorphism
problem is of practical importance in applications such as storing and
retrieving molecular structure data from a
database\cite{Kuramochi-Karypis06} or verification of printed circuit
layout with respect to a schematic\cite{Ebeling-Zajicek83}.  It is
also interesting, because it is not known whether the problem in
general can be solved in polynomial time, or whether it is
$NP$-complete.
  
A {\em graph automorphism} is an isomorphism of a graph onto itself.
The set of graph automorphisms of a graph forms a group under
composition. The graph automorphism problem is the problem of finding
whether a graph has any automorphism other than the identity
automorphism, which like the graph isomorphism problem has unknown
computational complexity\cite{Lubiw81}. More generally, one is interested in the
size of the automorphism group\cite{Standish05a}, and the orbits of the group.

The graph isomorphism problem can be reduced to the problem of finding
a canonical labeling of the vertices of a graph. If the adjacency
matrices of two graphs under their canonical labeling are equal, then
they are isomorphic. For many graphs, finding a canonical labeling is
tractable, and Nauty\cite{McKay81} is probably considered one of the
best-of-breed implementations. Nauty will also return the size of the
automorphism group of a graph.

Unfortunately, Nauty struggles with star-like graphs, ie graphs where
several isomorphic graphs are attached to each other via a single hub
vertex. In this paper, I present the SuperNOVA, or ``star exploder''
algorithm, which can handle these sorts of graphs efficiently.

\section{The algorithm}

\subsection{Canonical Ranges}\label{canonRange}

The algorithm proceeds by defining an ordering relation on the graph
vertices, sorting the vertices according to that ordering relation and
the assigning a range of possible canonical labels to each vertex
according to its position in the sorted list. For example, if the
following sorted list was returned:
\begin{displaymath}
n_3 < n_2 = n_4 = n_5 < n_1 = n_0
\end{displaymath}
then the vector of canonical ranges will look like
\begin{displaymath}
[4,5), [4,5), [1,4), [0,1), [1,4), [1,4).
\end{displaymath}

\begin{enumerate}
\item Initially, the ordering relation used is vertex degree
  (both in-degree and out-degree, and the number of bidirectional
  edges). 
  
\item Once all vertices have been assigned a canonical range, we can
  compare the canonical ranges of the nearest neighbours of a pair of
  vertices. If two vertices have the same canonical range, yet their
  neighbourhoods differ, we can further discriminate between the
  vertices, enabling a refinement of the canonical ranges.  This step
  is repeated until no further refinement is possible.
\end{enumerate}

The computational complexity lies between $O(n\log n)$ (the
computational complexity of a sort) and $O(n^2\log n)$ as at most $n$
iterations can occur in step 2.

If the result of this algorithm is that every vertex has a canonical
range of size 1, then we are done. The canonical labeling is given by
the lower bounds of the canonical ranges, and there is only one
automorphism (the identity). However, if some of the vertices have
non-unit ranges, then the graph may have symmetries. Unfortunately, we
cannot just take the product of the ranges as the size of the
automorphism group, as not all such relabelings are automorphisms.

\subsection{Symmetry Breaker}

If the graph has symmetries, then at least two vertices will have
identical canonical ranges. We need to determine which of the possible
labelings is a canonical labeling. There may be more than one
canonical labeling, but each such labeling produces an identical
adjacency matrix. To compute a canonical labeling, we induce an
ordering over adjacency matrices, and pick a labeling having the
least adjacency matrix.

The outline of the symmetry breaker algorithm is:
\begin{tabbing}
XX \= XX \= XX \= \kill
If all canonical ranges are of size 1, then \\
\> return an automorphism count of 1, and \\
\>\> an adjacency matrix for that labeling,\\
otherwise \\
Find first non-unit canonical range $[m,M)$\\
\> For  each  vertex $j$ having canonical range $[m,M)$, \\
\>\> set vertex $j$'s canonical range to $[m,m+1)$\\
\>\> apply step 2 of \S\ref{canonRange}\\
\>\> recursively apply the symmetry breaker algorithm to the new\\
\>\>\> canonical ranges.\\
\>\> add the returned automorphism count to the map entry indexed by\\
\>\>\>the returned adjacency matrix\\
\>return the least adjacency matrix and its automorphism count\\
\end{tabbing}

The worst case scenario for this algorithm is when the sorting
algorithm in \S\ref{canonRange} fails to discriminate vertices, in which
case the complexity is $O(n!)$, as each permutation of vertices will be
tried by the symmetry breaker. This will occur for the fully connected
graph, which will always be a worst case, but also for the empty graph
and star configurations. A star graph of order $n+1$ containing a
single hub of degree $n$, and $n$ leaf vertices, will cause the symmetry
breaker algorithm to have complexity $O(n!)$. Given that this is
the same problem that afflicts Nauty, this leads naturally to the star
exploder algorithm.

\subsection{Star Exploder}

If we have a simple star topology, with $c$ isomorphic graphs attached
to a central hub, then the automorphism group size is given by $c!r^c$,
where $r$ is the automorphism group size of each of the spokes of the
star, since there are $c!$ ways of relabeling the spokes. A
slightly more general case occurs where there are $c_0$ spokes isomorphic to
each other, another group of $c_1$ spokes isomorphic to each other and
so on. In this case, the resulting automorphism group size is given by
\begin{equation}\label{stareqn}
r=\prod_ic_i!r_i^{c_i}.
\end{equation} 

To establish whether an arbitrary graph has a star-like topology, we
remove all vertices with unit canonical range, which we call ``fixed
vertices''. A graph colouring algorithm can be used to find the
different contiguous subgraphs. If the graph breaks into more than one
contiguous piece, then we can recursively apply the complete
automorphism algorithm to each piece to obtain $r_i$, and count each
piece using a map indexed by the canonical adjacency matrix. Then the
overall automorphism group size can be found from the individual size
by using equation (\ref{stareqn}). The overall canonical labeling can
be found by using the algorithm described in \S\ref{canonRange}, but
with a modified ordering that includes information about which
subgraph the vertices belong to (subgraphs sorted according to their
canonical adjacency matrix order), and the canonical label of the vertex
within the subgraph. If two vertices belong to different, but
isomorphic subgraphs, and further that they have the same canonical
label within their respective subgraph, then they are ordered simply
by their original label, as in this case it wouldn't matter which way
they were labeled, the adjacency matrix would be identical. This
allows a canonical labelling to be generated.

A subtle twist to be considered here is that a subgraph connected
to one fixed vertex, and another subgraph connected to a different fixed
vertex are not equivalent, even thought they may be isomorphic. To deal
with this issue, we attach the vertex's canonical range from the
original graph as an attribute to the equivalent vertex in the
subgraph. Only isomorphic subgraphs whose attributes are identical are
equivalent, otherwise they're counted as distinct graphs. 

Star exploder fails when either there are no fixed vertices, or when
removing all the fixed vertices does not partition the graph.  Because
the symmetry breaker algorithm gradually fixes more and more vertices at
each level of recursion, the star exploder algorithm is applied at
each level of recursion of the symmetry breaker algorithm, and will
eventually succeed in breaking the graph into disjoint pieces. The
worst case scenario is not so much the full graph (which being the
dual of the empty graph is trivially transformed), but a digraph where
each vertex is connected to every other vertex, and arranged so that
the indegree and outdegree of each vertex is the same (the order of
the graph must be odd for this to occur). Each vertex is the same as
any other, so symmetry breaker must iterate over all $n!$
permutations of vertex labels.

\section{Implementation and results}

The algorithm was implemented in C++ as part of the open-source
\EcoLab{}\cite{Standish-Leow03} simulation environment, from
ecolab.4.D31. onwards. It makes use of the C++ standard library sort()
algorithm, and the standard associative containers map and set. 

The algorithm was tested by comparing its calculated automorphism
group size with that given by Nauty. If $S(g)$ is the canonical
representation of $g$ calculated by SuperNOVA, and $N(g)$ the
canonical representation calculated by Nauty, a second important check
is that $S(N(g))=S(g)$ and that $N(S(g))=N(g)$.

A database of 48940 symmetric graphs obtained from Brendan McKay's
website (http://cs.anu.edu.au/\~{}bdm/data/graphs.html) was used to
check the equivalence of SuperNOVA with Nauty. Exhaustively generating
all digraphs of a certain number of vertices and edges provided an
independent test to ensure the algorithm worked for digraphs,
including some with a star-like nature, however this was only feasible
up to order 9 or so. Certain star-like digraphs extracted from the
wiring diagram of a {\em C. elegans} brain was used to test the
performance of SuperNOVA on graphs that proved intractable with Nauty.
Attempting to run Nauty on these digraphs was unsuccessful, as Nauty
didn't complete after several days of running, and had to be killed.
By contrast, SuperNOVA computed these examples in seconds.

\begin{figure}
\epsfxsize=\textwidth
\epsfbox{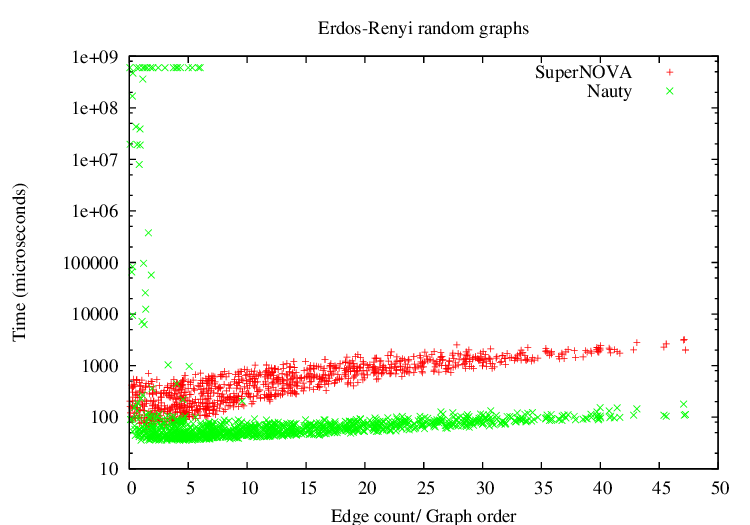}
\caption{Execution times for randomly generated Erd\"os-R\'enyi graphs
  with order $10\le n<100$ and edge count $0<l\le n(n-1)/2$ by
  SuperNOVA and Nauty. Times are plotted against $l/n$. Whilst for
  most graphs, Nauty is an order of magnitude faster, for a number of
  graphs it is many, many times slower. These cases are all for low
  order/edge count ratios ($<7$).}
\label{randomGraph}
\end{figure}

Figure \ref{randomGraph} shows 1000 randomly generated Erd\"os-R\'enyi
graphs with order $10\le n<100$ and edge count $0<l\le n(n-1)/2$. Both
SuperNOVA and Nauty were timed, and the times plotted as a function of
order and edge count. Because some graphs can potentially take a very
long to compute the canonical labeling, a maximum of 10 minutes was
imposed on the computation by using CPU resource limit functionality
of Linux. These examples appear in the data as having an execution
time of 10 minutes ($6\times10^8\mu $s), and all from Nauty, and
appear when $l/n<10$.

\begin{figure}\
\epsfxsize=\textwidth
\epsfbox{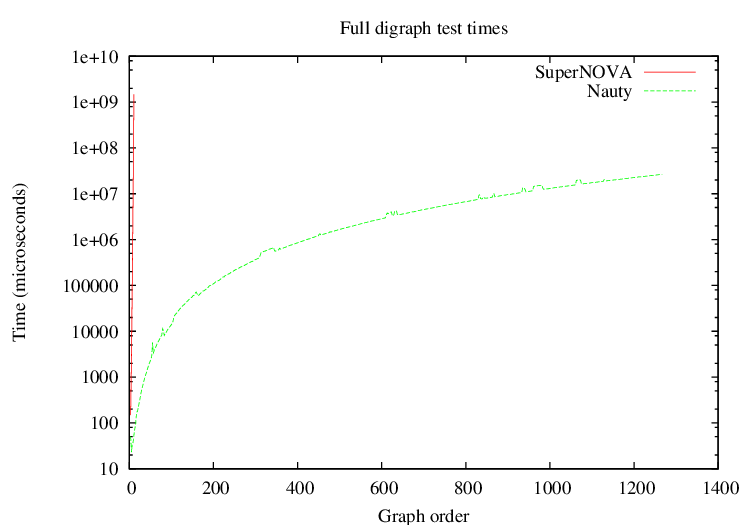}
\caption{Execution times for the full digraph case as a function of
  graph order. This is the worst case for SuperNOVA, which shows super
  exponential complexity. Nauty exhibits polynomial complexity}
\label{worstCase}
\end{figure}

Figure \ref{worstCase} shows the performance of SuperNOVA versus Nauty
for the worst case scenario of a fully connected digraph, with the
indegree and outdegree of each vertex being the same. As expected,
SuperNOVA's execution time blows up very rapidly, but interestingly
Nauty performs well, with polynomial time complexity.

\section{Discussion}

SuperNOVA effectively handles the star-like configurations that give
trouble to Nauty. On a sampling of 1000 Erd\"os-R\'enyi random graphs
of order between 10 and 100, SuperNOVA computed the automorphism group
size of all of the whole set within 10 minutes on a quad core Intel
Core 2. By contrast, Nauty failed to complete the calculations on
several of the graphs within the 10 minute time limit. Interestingly,
all of these examples lie in the range $0<l/n<10$. Overall, Nauty is
an order of magnitude faster than SuperNOVA on those examples it
handles well, which is a reflection of the intense effort that has
goine into the optimisation of that code. With more optimisation,
SuperNOVA should be able to close the gap somewhat.

On the artificially constructed worst-case scenario, SuperNOVA
performs poorly as expected, but Nauty performs well, executing in
polynomial time.

All of this suggests the possibility of a hybrid algorithm, leaving
the sparse examples to SuperNOVA, and the denser examples to Nauty. The
precise heuristic for determining which algorithm will need to be
determined by future research. Furthermore, the precise canonical form
returned differs in the two algorithms, so combining the algorithms will
need to take this into account. For the isomorphism problem, it should
not matter, so long as the same algorithm (SuperNOVA or Nauty) is
applied to the two graphs being compared.

Finally, one may speculate as to whether the hybrid algorithm is truly
polynomial complexity. Further work will be needed to try and identify
worst case scenarios for both algorithms.

\section{Note added after review}

Reviewers of this paper thought the algorithms described here are
well-known. The first reviewers comment is:
\begin{quote}

This paper describes an algorithm for graph isomorphism/automorphism,
and makes some comparisons with the program ``nauty''.

It seems to me that the author has just rediscovered the known
problem that nauty has with directed graphs, which can be solved
using the invariant ``adjacencies'' as described in the nauty manual.
This entirely cures the anomalies noted in Fig 1 and then nauty is
consistently more than one order of magnitude faster than SuperNOVA
for these easy digraphs.

Whether this explains the problem with ``star topologies'' as well,
it is hard to tell since not a single example is given. In addition,
it is hard to tell since not a single example is given. In addition,
no reason is offered for why this class of graphs is significant.

No effort is made to track down truly difficult graphs for comparison.
This is a serious deficiency, because what is here is much too simple
to have serious chance of competing.

The processes of ``canonical ranges'' and ``symmetry breaker'' are not
new, but are just the standard approaches of partition refinement
and vertex individualisation that most successful isomorphism
algorithms (including nauty) have used for decades.

The formula $c!r$ should be $c! r^c$, and similarly for
(\ref{stareqn}).  Also, when used properly nauty does not exhibit the
n! behaviour which this paper claims it does.  \end{quote}

With respect to the last comment, the referee is correct, the paper is
in error, but the implementation gets it correct. I have therefore
adjusted this paper.

I accept that the canonical range and symmetry breaker algorithm are
well known, and go by the names of ``vertex refinement'' and
``individualisation'', as pointed out by the second reviewer. I cannot
include the second reviewer's review inline here, as it was delivered
via PDF. It is available via my
website\cite{automorph-review2}. However, the key ``SuperNOVA'', or
``Star Breaker'' algorithm, is not so well known. The second reviewer
missed this innovation completely. As I will explain in the following
section ``Algorithm Race'', even with the adjacency invariance option
of Nauty enabled, it is not competitive with SuperNOVA. Nevertheless,
I also tested two other packages which have since come to my
attention: Saucy\cite{Darga-etal08} and the igraph package, which includes the
VF2 algorithm\cite{Cordella-eta01}, and the BLISS
algorithm\cite{Junttila-Kaski07}. Saucy performs better than Nauty on
the problematic graphs, but is still not competitive aith
SuperNOVA. However BLISS appears to manage those graphs, particularly
for larger orders.

\subsection{Algorithm Race}

To assess the validity of the various algorithms, and their
implementations, I concocted an experiment called an ``algorithm
race''. 

Because graph automorphism algorithms are combinatorically expensive
in the worst case, the usual technique of benchmarking programs is
impractical. However, all we're really interested in is which
algorithm solves the problem fastest. So the idea is to {\em fork} the
computation, and run a separate algorithm on each process. When the
first algorithm finishes, it posts the result to the master process
via a {\em pipe}. The master process then kills all the processes. In
order to ensure the computation always terminates, one of the
``algorithms'' simply sleeps for a given amount of time. We note these
cases as ``timed out''. On a multicore system, all processes get equal
priority, provided the number of algorithms tested is less than or
equal to the number of cores.

This cannot be done with usual threading models, as, for example,
posix threads do not supply a way of killing threads. The posix
fork/join and pipe model provides the necessary functionality. This
still works in Windows when the code is compiled using the Cygwin
posix emulation layer, but not the Mingw, which just wraps the
Windows API.

The experiment involved creating sparse Erd\"os-R\'enyi random graphs,
where the number of links is no more than 10 times the graph order, as
sparse graphs are the only graphs where SupeNOVA has an advantage. The
first experiment involved graphs of order 5000, and a timeout of 60
seconds, and the second experiment of order 20,000. The code is
implemented in C++ (\EcoLab{} version 5.D7), and was run on an 8 core
AMD FX 8150 at 3.6GHz.

\begin{table}
\begin{tabular}{lrr}
Algorithm & Wins for 5000 nodes & Wins for 20,000 nodes\\\hline
Timed out & 0 & 15 \\
SuperNOVA & 4125 & 0 \\
Nauty & 3997 & 52 \\
Nauty (adj inv) & 1072 & 11 \\
Saucy & 385 & 7 \\
VF2 & 1 & 0\\
Bliss(0) & 9578 & 733 \\
Bliss(1) & 2699 & 52 \\
Bliss(2) & 5198 & 500 \\ \hline
Total & 27046 & 1370 \\\hline
\end{tabular}
\caption{Results of the algorithm race for random graphs of order 5000
and order 20000. The number after BLISS refers to the splitting
heuristic (0=first non-singleton call, 1=first largest non-singleton
call, 2=first smallest non-singleton call).}
\label{algoWinners}
\end{table}

From Table \ref{algoWinners}, we can conclude that the 1st and thris
variant of BLISS did best overall, particularly for the larger graphs,
the two variants of Nauty also did well, as well as SuperNOVA. So a
hybrid algorithm, based on the race idea with these 5 algorithms, is
what is now implemented in EcoLab. If igraph is not available, the
only the two nauty variants and SuperNOVA will be used.

Another experiment involved saving the SuperNOVA winning graphs from
the order 5000 experiment, and then rerunning the race with SuperNOVA
absent. In this case there were many timeouts after 60 seconds, but
all completed with 10 minutes. If the Bliss algorithm was also
excluded, then there were many timeouts after 10 minutes, indicating
that none of Nauty, Saucy or VF2 were effective in these cases. In
particular, Nauty could not have implemented SuperNOVA. However, Bliss
is effective, even on the SuperNOVA winning cases. This could mean
that Bliss has implemented the SuperNOVA algorithm, or at very least
an algorithm that is as effective.


\end{document}